\documentclass[letterpaper, 10 pt, conference]{ieeeconf}  

\IEEEoverridecommandlockouts                              
\usepackage[utf8]{inputenc}
\usepackage[T1]{fontenc}
\usepackage{graphicx}
\graphicspath{ {./images/} }

\usepackage{sgame}
\usepackage{amssymb}
\usepackage{amsmath}
\usepackage{stmaryrd}
\usepackage{xcolor}
\usepackage{float}
\usepackage{comment}
\usepackage{lipsum}
\usepackage{mathbbol}
\usepackage{hyperref}
\usepackage[capitalize]{cleveref}
\usepackage{enumerate}
\usepackage{cite}
\usepackage{mathtools}
\usepackage{amssymb}
\usepackage{tikz}

\newenvironment{proofsketch}{\par\noindent\textit{Proof Sketch:}}{\hfill$\square$\\}

\newtheorem{remark}{Remark}
\newtheorem{proposition}{Proposition}
\newtheorem{theorem}{Theorem}

\newtheorem{definition}{Definition}

\DeclareMathOperator{\Equaldef}{\overset{def}{=}}

\title{\LARGE \bf Optimal Teaming for Coordination with Bounded Rationality via Convex Optimization}

\author{Zhewei Wang, Olugbenga Moses Anubi and Marcos M. Vasconcelos   
\thanks{Z. Wang is with the Department of Mechanical Engineering, FAMU-FSU College of Engineering, Florida State University,  Tallahassee, FL 32306, USA. O. M. Anubi and M. M. Vasconcelos are with the Department of Electrical and Computer Engineering, FAMU-FSU College of Engineering, Florida State University,  Tallahassee, FL 32306, USA. E-mails:
        {\tt   zw23a@fsu.edu, oanubi@fsu.edu, m.vasconcelos@fsu.edu}.}%
}

\begin{document}

\maketitle
\thispagestyle{empty}
\pagestyle{empty}

\begin{abstract}
Teaming is the process of establishing connections among agents within a system to enable collaboration toward achieving a collective goal. This paper examines teaming in the context of a network of agents learning to coordinate with bounded rationality. In our framework, the team structure is represented via a weighted graph, and the agents use log-linear learning. We formulate the design of the graph's weight matrix as a convex optimization problem whose objective is to maximize the probability of learning a Nash equilibrium while minimizing a connectivity cost. Despite its convexity, solving this optimization problem is computationally challenging, as the objective function involves the summation over the action profile space, which  grows exponentially with the number of agents. Leveraging the underlying symmetry and convexity properties of the problem, when there are no sparsity constraints, we prove that there exists an optimal solution corresponding to a uniformly weighted graph, simplifying to a one-dimensional convex optimization problem. Additionally, we show that the optimal weight decreases monotonically with the agent's rationality, implying that when the agents become more rational the optimal team requires less connectivity. 
\end{abstract}

\section{Introduction}

In strategic decision-making over a social network, every agent must take into account the behavior of its neighbors when choosing an action. Agents in the system are often characterized by a certain level of rationality, which can be modeled on how much weight (if any) is assigned to the actions taken by its neighbors. In one extreme, a completely irrational agent makes decisions by randomly picking an action over a finite discrete set with a uniform distribution. On the opposite extreme, a fully rational agent optimizes its local objective taking into account the actions played by its neighbors, in other words, the agent plays a \textit{best-response}. Somewhere in between these extremes, an agent with bounded rationality will make suboptimal decisions with non-negligible probability, resulting in loss of performance.

The focus of  this paper is on the design of undirected weighted networks when the agents are coordinating binary actions under bounded rationality using log-linear learning (LLL).
The objective of the system designer consists of two terms: the first is the stationary probability that the agents will learn to coordinate at the game's Nash equilibrium and the second is a cost on the overall connectivity of the graph, represented here by the total sum of the graph weights. We show that the optimal design of weights for any specified connectivity/sparsity pattern is a convex optimization problem. Then, we establish that if the sparsity pattern is complete, there exists an optimal graph with uniform weight distribution over the edges. The optimal weight is a non-trivial function of the agents' rationality parameter, the number of agents in the graph, and the connectivity cost. We analytically show that as the agents become more rational, the optimal weights decrease and vice-versa. To the best of our knowledge this is the first time such results appear in the literature. We also report numerical results that show that when incomplete sparsity pattern constraints are introduced, the optimal solutions are asymmetric in general. 


\begin{figure}[t!]
    \centering
\includegraphics[width=\columnwidth]
{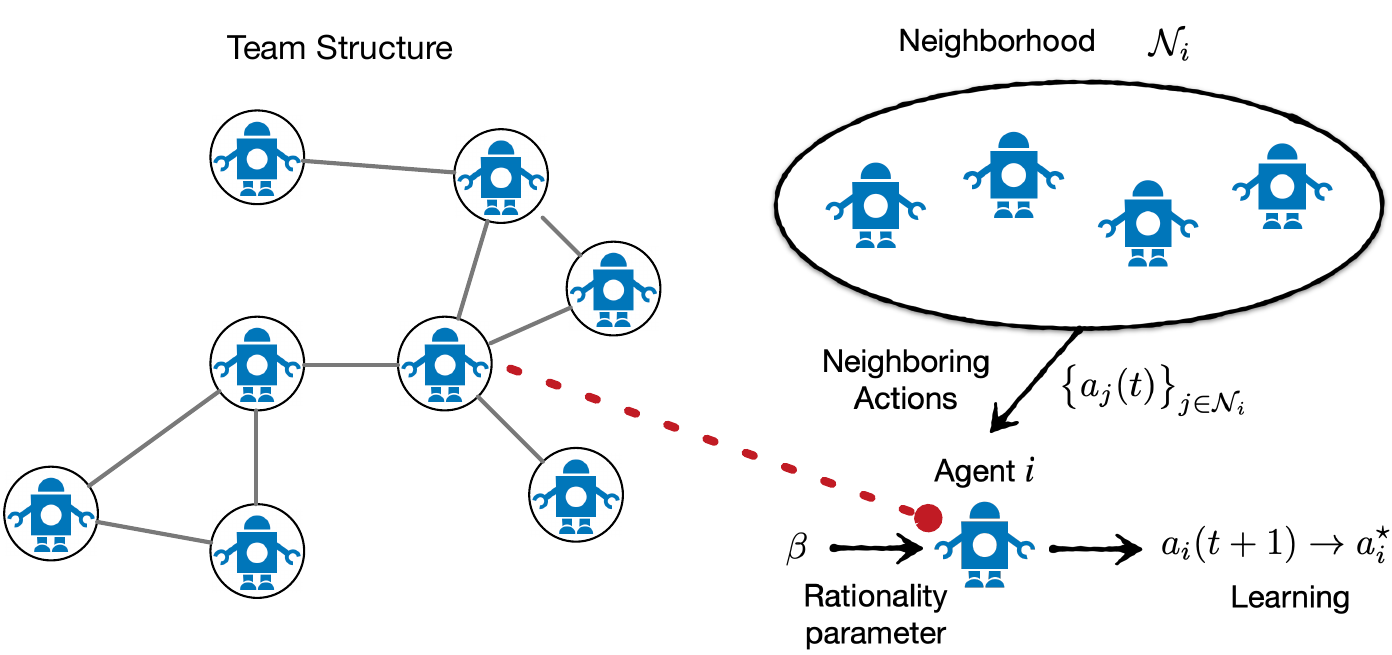}
    \caption{Framework for learning to coordinate over a network with bounded rationality. The teaming problem is to choose the weights that maximize the probability of learning to coordinate at a Nash equilibrium.}
    \label{fig:LLL}
\end{figure}

\subsection{Related Literature}

The interplay between network structure and agent behavior in coordination games has been extensively studied. Foundational contributions such as \cite[and references therein]{jackson2015games,9833086} introduced general network game models. The seminal work reported in \cite{blume1993statistical} established the first analytical results in the Markovian dynamics and stationary distributions of LLL. Additional convergence properties of LLL, particularly under asynchronous updates, were further analyzed in \cite{marden2012revisiting}. Recent studies, such as \cite{paarporn2020risk,paarporn2020impact}, applied LLL to network coordination games in an adversarial setting.

One of the important questions in the field of network games is how the network topology influences coordination efficiency, and the convergence rate of the learning process. The studies in \cite{Montanari:2010,arieli2020speed} shows how structural properties affect the speed and likelihood of converging to the right Nash equilibrium. However, the afore mentioned results rely on the agents' rationality to grow unbounded. We are interested in the case where the agents operate under bounded rationality.

Due to cognitive, informational and computational limitations, the concept of bounded rationality has gained increasing attention in recent literature \cite{guan2021bounded,tsiotras2021bounded,kokolakis2023bounded,paarporn2024madness}. 
Learning for coordination in the presence incomplete information have been considered in  \cite{Eksin:2013,Wei:2023,Vasconcelos:2023}. Adaptive learning algorithms under bounded rationality have also been explored in \cite{chasnov2024dynamics,xianjia2023rationality}, where the agents' rationality strongly influences their ability to coordinate, adaptation speed, and long-term performance. Recent work in \cite{9482895} captures bounded rationality by assuming that the agents only have local information about constraints in a network game.
Our prior work \cite{zhang2024role,zhang2024rationality} showed that increased connectivity improves coordination under bounded rationality in LLL, both in regular and irregular graph structures.

Motivated by \cite{10453658}, we considered the extension of our prior work to the design and analysis of weighted graphs for learning to coordinate under bounded rationality. Our results reported herein use convex optimization techniques in a different context than the now classic results from \cite{boyd2006convex}, and provide many new research directions in the design of network systems of strategic agents.

\subsection{Notation}
Let \( N \) denote the number of agents in the network. Each agent selects an action \( a_i \in \{0,1\} \), and the joint action profile is denoted by \( a = [a_1, \ldots, a_N]^T \in \{0,1\}^N \). The communication topology is modeled as an undirected graph with symmetric weight matrix \( \mathbf{W} = [w_{ij}] \in \mathbb{R}^{N \times N} \), where \( w_{ij} \in [0,1] \) represents the weight of the edge between agents \( i \) and \( j \), and \( w_{ij} = 0 \) if \( (i,j) \notin \mathcal{E} \).

The scalar \( \theta \in \mathbb{R} \) is a  parameter representing task difficulty. The parameter \( \beta \geq 0 \) represents the agents’ rationality level. The function \( \Phi_\mathbf{W}(a) \) denotes the potential function evaluated at action profile \( a \), and \( \mu_\mathbf{W}(a \mid \beta) \) is the stationary distribution over joint actions under log-linear learning with weight matrix \( W \) and rationality \( \beta \). The regularization parameter \( \rho > 0 \) penalizes total connectivity in the graph. The vectors \( \mathbb{1},\mathbb{0} \in \mathbb{R}^N \) denote the all-ones and the all-zeroes column vectors. The identity matrix is denoted by $\mathbf{I}$.

\subsection{Organization}
The remainder of this paper is organized as follows. Section II introduces the networked coordination game setup, defines the potential function, and formulates the weight optimization problem. In Section III, we analyze the structure of the optimal weight matrix. In Section IV we characterize the dependence of the optimal graph weight with the rationality parameter. Section V extends the formulation to graphs with arbitrary sparsity patterns and discusses implications for network design in the face of bounded rationality. In Section VI, we conclude the paper.

\section{Weighted Network Coordination Games}

\subsection{Definition and preliminary properties}

Consider a team with $N$ agents connected by a weighted graph $\mathcal{G}([N],\mathcal{E})$, i.e., each link in the graph is is associated with a weight $w_{ij} \in [0,1]$. If $(i,j)\notin \mathcal{E}$, then $w_{ij} = 0$. Otherwise, $w_{ij}>0$. The team structure is characterized by a weight matrix $\mathbf{W}$, whose entry $w_{ij}$ represents the weight between agents $i$ and $j$.

Our framework consists of a network coordination game defined on the weighted graph described by $\mathbf{W}$. 
Consider a binary coordination game played between two agents in the system. Let $(i,j)\in\mathcal{E}$, and suppose that $a_i,a_j\in\{0,1\}$ are the actions played by agents $i$ and $j$, respectively. The bimatrix game in \cref{fig:bimatrix} specifies the payoffs for the pairwise interaction between $i$ and $j$, where the parameter $\theta \in \mathbb{R}$, represents the underlying difficulty of performing a collective task (c.f. \cref{rmk:task_allocation}).
\begin{figure}[h!]\hspace*{\fill}%
\centering
\begin{game}{2}{2}[$a_i$][$a_j$]
     & $1$ & $0$             \\
 $1$ & $\Big(1-\frac{\theta}{N},1-\frac{\theta}{N}\Big)$ & $\Big(-\frac{\theta}{N},0\Big)$ \\
 $0$ & $\Big(0,-\frac{\theta}{N}\Big)$ & $\Big(0,0\Big)$  \\
\end{game}\hspace*{\fill}%
\caption{A coordination game with parameter $\theta$ between two players.}
\label{fig:bimatrix}
\end{figure}

\begin{remark}[Payoff Interpretation]\label{rmk:task_allocation}
One of the applications of the framework proposed herein is in distributed multi-robot task allocation \cite{Wei:2023,Vasconcelos:2023,Kanakia:2016}. In such application, there is a single task of difficulty $\theta$, which is amortized among $N$ agents in the system forming subtasks of difficulty $\theta/N$. Each agent has one unit of power to use towards completing its subtask, but will only succeed if it works together with its neighbors. Hence, the coordination game in \cref{fig:bimatrix}, which is an instance of a \textit{Stag-Hunt} game \cite{Fudenberg:1998}.
\end{remark}

\vspace{5pt}

Extending the two-agent coordination game from \cref{fig:bimatrix} over a network, we obtain the following  \textit{network game} with $N$ agents, where agent $i$ simultaneously plays the same action with all of its neighbors $j\in\mathcal{N}_i$. Let $V_{ij}:\{0,1\}^2\rightarrow \mathbb{R}$ be defined as:
\begin{equation}
V_{ij}(a_i,a_j) \Equaldef a_i\Big(
a_j-\frac{\theta}{N}\Big).
\end{equation}

In a network coordination game, every agent receives the weighted sum of all the payoffs of the bimatrix games $V_{ij}(a_{i}, a_{j})$  played with each of its neighbors. Therefore, for the $i$-th agent, the utility is determined as follows
\begin{equation}\label{eq:shapley}
    U_{i}(a_{i},a_{-i})\Equaldef \sum\limits_{j\in \mathcal{N}_i} w_{ij} V_{ij}(a_{i},a_{j}).
\end{equation}
Therefore, the payoff of the $i$-th agent in our game is
\begin{equation}\label{eq:payoff_network}
U_{i}(a_{i},a_{-i}) = a_i\Big(\sum_{j\in \mathcal{N}_i}w_{ij}a_j-\frac{\theta}{N}\sum_{j \in \mathcal{N}_i}w_{ij}\Big).
\end{equation}
Notice that the payoff in \cref{eq:payoff_network} reflects that more connected node face a larger difficulty due to the collaboration with other agents in the network. However, the effort is also aggregated, which leads to a potentially higher utility.

\vspace{5pt}

\begin{definition}[Exact potential games]\label{exact_potential}
Let $\mathcal{A}_i$ denote the action set of the $i$-th agent in a game with payoff functions $U_i(a_i,a_{-i})$, $i\in[N]$. Let $\mathcal{A} = \mathcal{A}_1\times \cdots \times \mathcal{A}_n$. A game is an \textit{exact potential game} if there exists a \textit{potential function} {$\Phi$}: 
 {$\mathcal{A}\rightarrow \mathbb{R}$} such that 
\begin{equation}\label{eq:exact_potential}
    U_{i}(a'_{i},a_{-i})- U_{i}(a''_{i},a_{-i}) = \Phi(a'_{i},a_{-i})- \Phi(a''_{i},a_{-i}),
\end{equation}
for all  $a_{i}',a_{i}''\in \mathcal{A}_{i}$, $a_{-i}\in \mathcal{A}_{-i}$,  $i \in [N]$.
\end{definition}

\vspace{5pt}

\begin{proposition}
The network coordination game defined above is an exact potential game if and only if the graph's weight matrix $\mathbf{W}$ is symmetric, i.e., $w_{ij} = w_{ji}$ for all $i,j$.
\end{proposition}

\vspace{5pt}

\begin{proofsketch}
From Monderer and Shapley \cite[Theorem 4.5]{monderer1996potential}, we know that  when a game has continuous action sets, which are intervals of real numbers, and the payoff functions are twice continuously differentiable, it is a potential game if and only if 
\begin{equation}
\frac{\partial^2 U_i}{\partial a_i \partial a_j} = \frac{\partial^2 U_j}{\partial a_i \partial a_j}, \quad i, j \in[N].
\end{equation}

Relaxing the action set $\mathcal{A}_i=\{0,1\}$ to $\tilde{\mathcal{A}}_i=\mathbb{R}$.   
Since the payoffs are twice continuously differentiable, the condition in \cref{eq:shapley} becomes
\begin{equation}
w_{ij} = w_{ji}, \ i,j \in [N] \Rightarrow \mathbf{W} = \mathbf{W}^\mathsf{T}.
\end{equation}
Since this condition must hold for the relaxed action set $\tilde{\mathcal{A}}_i$, it must also hold for $\mathcal{A}_i\subseteq \tilde{\mathcal{A}}_i$.  
\end{proofsketch}

\vspace{5pt}

To obtain a potential function for our network game, define the following function
\begin{equation}\label{eq:potential_two_agents}
\phi_{ij} (a_i,a_j) \Equaldef   a_ia_j + (1-a_i-a_j)\frac{\theta}{N}. 
\end{equation}
The function $\phi_{ij}$ is a potential function for the (weighted) two-agent coordination game in \cref{fig:bimatrix}, i.e., 
\begin{equation}
\phi_{ij} (1,a_j) -\phi_{ij} (0,a_j)  = V_{ij}(1,a_j) - V_{ij}(0,a_j),  
\end{equation}
for all $i,j\in [N]$ and $a_j \in \{0,1\}$. Let $\Phi_{\mathbf{W}}: \{0,1\}^N \rightarrow \mathbb{R}$ as follows
\begin{equation}\label{potential_function}
    \Phi_{\mathbf{W}}(a) \Equaldef \frac{1}{2}\sum\limits_{i\in [N]} \sum\limits_{j\neq i}w_{ij} \phi_{ij}(a_{i},a_{j}).
\end{equation}

Evaluating the summations, we obtain the potential function $\Phi_\mathbf{W}$ for the network coordination game with weight matrix $\mathbf{W}$ as
\begin{equation}
 \Phi_{\mathbf{W}}(a) = \frac{1}{2}a^\mathsf{T}\mathbf{W} a - \frac{\theta}{N}\mathbb{1}^\mathsf{T}\mathbf{W}a + \frac{\theta}{2N}\mathbb{1}^\mathsf{T}\mathbf{W}\mathbb{1}.\textbf{}
\end{equation}
 
\vspace{5pt}

Note that, while $\Phi_\mathbf{W}(a)$ is a binary quadratic form in $a$, it is a linear function of $\mathbf{W}$. We state this property formally, without proof, in the following proposition.

\vspace{5pt}

\begin{proposition}
For a fixed $a \in \{0,1\}^N$, the potential function $\Phi_{\mathbf{W}}(a)$ is linear in $\mathbf{W}$.
\end{proposition}

\vspace{5pt}

A fundamental result from Shapley and Monderer \cite{monderer1996potential}, establishes a connection between the Nash Equilibria of the network coordination game and the maximizers of the corresponding potential function. Rewriting  
$\Phi_\mathbf{W}(a)$ in terms of the graph Laplacian, we obtain
\begin{equation}\label{eq:pot_laplacian}
 \Phi_{\mathbf{W}}(a) = -\frac{1}{2}a^\mathsf{T}\mathbf{L}(\mathbf{W}) a + \bigg(\frac{1}{2}- \frac{\theta}{N}\bigg)\mathbb{1}^\mathsf{T}\mathbf{W}a + \frac{\theta}{2N}\mathbb{1}^\mathsf{T}\mathbf{W}\mathbb{1},
\end{equation}
where $\mathbf{L}(\mathbf{W})\Equaldef \mathrm{diag}(\mathbf{W}\mathbb{1})-\mathbf{W}$. We characterize the set of Nash equilibria of the weighted network coordination game in the following proposition.

\vspace{5pt}

\begin{proposition}\label{prop:nash_equilibria}
For all $\mathbf{W}\in [0,1]^{N\times N}$, the following holds: If $\theta \leq N/2$, the potential function is maximized at $a^\star = \mathbb{1}$; If $\theta > N/2$, the potential function is maximized at $a^\star = \mathbb{0}$. Therefore, the set of Nash equilibria for the weighted network coordination game $\mathcal{S}_{\mathrm{NE}}$ satisfies
\begin{equation}
\mathcal{S}_{\mathrm{NE}}\subseteq \{\mathbb{0},\mathbb{1}\}.
\end{equation}
\end{proposition}

\vspace{5pt}

\begin{proof}
 The proof follows from \cref{eq:pot_laplacian} using the fact that $a^\mathsf{T}\mathbf{L}(\mathbf{W}) a = 0 \Leftrightarrow a\in\{\mathbb{0},\mathbb{1}\}$.
\end{proof}

\vspace{5pt}

\subsection{Log-linear learning}

Log-linear learning (LLL) is an algorithm that enables agents to converge to a Nash equilibrium in potential games. LLL is characterized by a parameter $\beta$, which controls the trade-off between exploration and exploitation, and can be interpreted as a measure of the agents' rationality. At time $t$, a randomly selected agent $i$ updates its action, given the actions of its neighbors, according to a softmax distribution, i.e.,
\begin{multline}
\mathbf{P}\big(A_i(t+1)=a_i \mid \{A_j(t)=a_j(t)\}_{j\in \mathcal{N}_i}\big)  \\ = 
\frac{ e^{\beta U_i\big(a_i,a_{-i}(t)\big)}}{e^{\beta U_i\big(0,a_{-i}(t)\big)}+e^{\beta U_i\big(1,a_{-i}(t)\big)}}, 
\ a_i\in \{0,1\}.
\end{multline}

When $\beta \rightarrow 0$, an agent will essentially choose an action from its action space uniformly at random. Conversely, as $\beta \rightarrow \infty$, the agent selects its next action according to a best response policy. Under bounded rationality (i.e., $\beta < \infty$), agents always have a nonzero probability of choosing a suboptimal action. LLL induces a Markov chain with a stationary distribution over the action space $\{0,1\}^N$ known as the Gibbs distribution, given by

\begin{equation}
\mu_\mathbf{W}(a \mid \beta) = \frac{e^{\beta \Phi_{\mathbf{W}}(a)}}{\sum_{a \in \{0,1\}^N}e^{\beta \Phi_{\mathbf{W}}(a)}}, \ \ a\in \{0,1\}^N.
\end{equation}

Using \cref{prop:nash_equilibria}, we can determine the probability of each of the possible equilibria for the game. If $a^{\star}=\mathbb{0}$ (when $\theta>N/2$), we have
\begin{equation}
\Phi_{\mathbf{W}}(\mathbb{0}) = \frac{\theta}{2N}\mathbb{1}^\mathsf{T}\mathbf{W}\mathbb{1},
\end{equation}
which implies that 
\begin{equation}\label{eq:case1}
\mu_\mathbf{W}(\mathbb{0} \mid \beta) = \frac{1}{\sum_{a \in \{0,1\}^N} e^{\beta \big(\frac{1}{2}a^\mathsf{T}\mathbf{W} a - \frac{\theta}{N}\mathbb{1}^\mathsf{T}\mathbf{W}a\big)}}.
\end{equation}
Similarly, if $a^{\star}=\mathbb{1}$ (when $\theta<N/2$), we have
\begin{equation}
\Phi_{\mathbf{W}}(\mathbb{1}) = \Big(\frac{1}{2}-\frac{\theta}{2N}\Big)\mathbb{1}^\mathsf{T}\mathbf{W}\mathbb{1},
\end{equation}
which implies that 

\begin{multline}
    \label{eq:case2}
\mu_\mathbf{W}(\mathbb{1} \mid \beta) = \\ \frac{1}{\sum_{a \in \{0,1\}^N}e^{\beta \big(\frac{1}{2}a^\mathsf{T}\mathbf{W} a - \frac{\theta}{N}\mathbb{1}^\mathsf{T}\mathbf{W}a+\big(\frac{\theta}{N}-\frac{1}{2}\big)\mathbb{1}^\mathsf{T}\mathbf{W}\mathbb{1}\big)}}.
\end{multline}

\subsection{Optimization problem}

The question we aim to answer is how to choose the graph weight matrix $\mathbf{W}$ such that the probabilities in \cref{eq:case1} or \cref{eq:case2} are maximized. 
We also impose a linear penalty on network connectivity. To that end, for $a^\star \in \{\mathbb{0},\mathbb{1}\}$ and $\rho>0$, we are interested in solving the following optimization problem
\begin{equation}\label{eq:opt_prob}
\begin{aligned}
\mathrm{minimize}  \quad & 
\frac{1}{\mu_\mathbf{W}(a^\star \mid \beta)}
+ \frac{\rho}{2}\mathbb{1}^\mathsf{T}\mathbf{W}\mathbb{1} \\
\text{subject to} \quad 
& \operatorname{trace}(\mathbf{W}) = 0 \\
& 0 \leq \mathbf{W} \leq 1 \\
& \mathbf{W} = \mathbf{W}^\mathsf{T} \\
& \lambda_2\big(\mathbf{L}(\mathbf{W})\big) > 0 
\end{aligned}
\end{equation}
with variable $\mathbf{W}\in \mathbb{R}^{N\times N}.$ 

\vspace{5pt}

\begin{remark}
The trace constraint in \cref{eq:opt_prob} ensures that the graph has no self-loops. The inequality constraints are understood component-wise. The symmetry constraint guarantees that the weighted network game is a potential game. Finally, the constraint on the algebraic connectivity $\lambda_2\big(\mathbf{L}(\mathbf{W})\big)$ ensures that the graph is connected, which in turn leads to a unique stationary distribution $\mu_\mathbf{W}(a \mid \beta)$.
\end{remark}

\section{Main results}

The optimization problem stated in the previous section is well-structured and admits a characterization of its optimal solution, presented in the following result.

\vspace{5pt}

\begin{theorem}\label{thm1}
The optimization problem in \cref{eq:opt_prob} is convex, and admits an optimal solution of the form
\begin{equation}
\mathbf{W}^\star = w^\star \big( \mathbb{1}\mathbb{1}^\mathsf{T} - \mathbf{I} \big),
\end{equation}
with $w^\star >0$.
\end{theorem}

\vspace{5pt}

\begin{proof}
Without loss of generality, assume $a^\star = \mathbb{0}$\footnote{The same structural result holds for $a^\star = \mathbb{1}$ and is omitted for brevity.}. The objective function in \cref{eq:opt_prob} is
\begin{equation}
f_0(\mathbf{W})\Equaldef \sum_{a \in \{0,1\}^N} e^{\beta \big(\frac{1}{2}a^\mathsf{T}\mathbf{W} a - \frac{\theta}{N}\mathbb{1}^\mathsf{T}\mathbf{W}a\big)}+\frac{\rho}{2}\mathbb{1}^\mathsf{T}\mathbf{W}\mathbb{1}.
\end{equation}

Each term in the sum over $\{0,1\}^N$ is a composition of an affine function of $\mathbf{W}$ with the convex and increasing function $\exp(\cdot)$, and is therefore convex. The first three constraints are linear in $\mathbf{W}$, and therefore convex. The last constraint involves the second smallest eigenvalue of the Laplacian matrix, which is a concave function of $\mathbf{W}$ \cite{boyd2006convex}. Therefore, the graph connectedness is also a convex constraint.

To prove the second part of the statement, notice that the feasible set is invariant to permutations in the following sense: Suppose $\mathbf{W}$ is feasible for the problem in \cref{eq:opt_prob}. Let $P$ be any permutation matrix in the set of all $\mathcal{P}$ permutation matrices of dimension $N$. Let $\mathbf{W}' = P\mathbf{W}P^\mathsf{T}$. The two matrices $\mathbf{W}$ and $\mathbf{W}'$ and their corresponding Laplacian matrices are similar, and therefore, have the same spectrum (same eigenvalues). Therefore, the algebraic connectivity constraint is preserved. The other three constraints are trivially satisfied by $\mathbf{W}'$. Therefore, the feasible set is invariant to permutations.

Since $f_0(\mathbf{W})$ is continuous and the feasible set $\mathcal{W}$ is compact, there exists a $\mathbf{W}^\star \in \mathcal{W}$ that achieves the minimum of $f_0(\mathbf{W})$, denoted by $f_0^\star$. Suppose  that $\mathbf{W}^\star$ is not of the form in the theorem statement. Based on $\mathbf{W}^\star$ we will construct an optimal solution to \cref{eq:opt_prob} of the form in the theorem statement with the same value $f_0^\star$. Consider $P \in \mathcal{P}$ and let $\mathbf{W}_P = P\mathbf{W}^\star P^\mathsf{T}$. Since $\mathbf{W}_P$ is feasible, we evaluate 
\begin{align*}
&f_0(\mathbf{W}_P)  =  \sum_{a \in \{0,1\}^N} \exp \left( \beta \left( \frac{1}{2} a^\mathsf{T} \mathbf{W}_P a - \frac{\theta}{N} \mathbb{1}^\mathsf{T} \mathbf{W}_P a \right) \right) \\
& =  \sum_{a \in \{0,1\}^N} \exp \left( \beta \left( \frac{1}{2} a^\mathsf{T} P\mathbf{W}^\star P^\mathsf{T} a - \frac{\theta}{N} \mathbb{1}^\mathsf{T} P\mathbf{W}^\star P^\mathsf{T} a \right) \right) \\
& \stackrel{(a)}{=}  \sum_{\tilde{a} \in \{0,1\}^N} \exp \left( \beta \left( \frac{1}{2} \tilde{a}^\mathsf{T}\mathbf{W}^\star \tilde{a} - \frac{\theta}{N} \mathbb{1}^\mathsf{T} \mathbf{W}^\star\tilde{a} \right) \right) \\
& =  f_0^\star,
\end{align*}
where $(a)$ follows from the change of variables $\tilde{a}=P^\mathsf{T}a$, and that the summation is over all possible vectors in $\{0,1\}^N$.

Define
\begin{equation}
\mathbf{W}_{\mathrm{sym}} \Equaldef \frac{1}{N!} \sum_{P\in \mathcal{P}} P\mathbf{W}^\star P^\mathsf{T}.
\end{equation}
Since $\mathbf{W}_{\mathrm{sym}}$ is feasible, and the objective function is convex, we have
\begin{multline}
f_0^\star \leq f_0(\mathbf{W}_{\mathrm{sym}}) = f_0\Big(\frac{1}{N!} \sum_{P\in \mathcal{P}} P\mathbf{W}^\star P^\mathsf{T} \Big) \\
\leq \frac{1}{N!} \sum_{P\in \mathcal{P}} \underbrace{f_0\big(\mathbf{W}_P\big)}_{=f_0^\star} = f_0^\star.
\end{multline}

Finally, we show that all the nonzero entries in $\mathbf{W}_{\mathrm{sym}}$ are equal. Since $\mathbf{W}_{\mathrm{sym}}$ is a symmetric matrix, without loss of generality, we can constrain our analysis to what happens when $\mathbf{W}^\star$ is an $N\times N$ strictly lower triangular matrix, such that
\begin{equation}
w_{ij}^\star = 0 \quad \text{for} \quad i \le j.
\end{equation}
Denote the subdiagonal entries of $\mathbf{W}^\star$ by
\begin{equation}
\{w^\star_1,w^\star_2,\ldots,w^\star_{M}\}\quad\text{where}\quad M = \frac{N(N-1)}{2}.
\end{equation}
Consider the set of all possible permutations $\Pi$ of these $M$ values. For $\pi \in \Pi$, we may construct a new matrix $\mathbf{W}_{\pi}^\star$ assigning them to the same strictly lower triangular positions. Let $\mathcal{P}$ denote the set of permutation matrices $P$ such that for any lower triangular matrix $\mathbf{W}$, the matrix $P\mathbf{W}P^\mathsf{T}$ is also lower triangular.
Therefore, for each permutation $\pi$, there exists  $P(\pi) \in \mathcal{P}$, such that $\mathbf{W}_{\pi}^\star =P(\pi)\mathbf{W}^\star P(\pi)^\mathsf{T}$. 

Therefore, the average of these matrices is
\begin{equation}
\mathbf{W}_{\text{sym}} =  \frac{1}{|\Pi|}\sum_{\pi \in \Pi} \mathbf{W}_{\pi}^\star = \frac{1}{|\mathcal{P}|}\sum_{P \in \mathcal{P}} P\mathbf{W}^\star P^\mathsf{T}.
\end{equation}
Since each nonzero position in $\mathbf{W}_{\pi}^\star$, across all permutations in $\Pi$, takes on each of the $M$ original subdiagonal values $w^\star_1, w^\star_2, \ldots, w^\star_M$ the same number of times, each such entry in $\mathbf{W}_{\text{sym}}$ is equal to
\begin{equation}
m = \frac{1}{M} \sum_{k=1}^{M} w^\star_k.
\end{equation}
\end{proof}

\section{Optimal Asymptotic Weight–Rationality Trade-off}

In the previous section, we have exploited the structure and convexity of the optimization problem in \cref{eq:opt_prob} to obtain a structural result to simplify the search for an optimal weight matrix. \Cref{thm1} implies that it is sufficient to search for a single weight $w^\star\in(0,1]$ that minimizes
\begin{equation}\label{optimal}
\tilde{f}_0(w) \Equaldef f_0\Big(w\big(\mathbb{1}\mathbb{1}^\mathsf{T} -\mathbf{I} \big)\Big).
\end{equation}

From now on, we will assume that $\theta > \frac{N}{2}$, which implies that $a^\star=\mathbb{0}$. We will show that there are non-trivial choices for $w$ as a function of $\beta$, $\rho$ and $N$. 
Evaluating $\tilde{f}_0(w)$, we get
\begin{multline}\label{eq:simplified_problem}
\tilde{f}_0(w)=\sum_{d=0}^N \binom{N}{d} \exp\bigg(\beta w \Big( \frac{d^2-d}{2}-\frac{\theta}{N}(N-1)d\Big)\bigg)\\
 + \frac{\rho}{2}wN(N-1).
\end{multline}

Here $d$ is the number of non-zero components in $a$, which simplifies the calculation when $\mathbf{W}$ is of the form in \cref{optimal}
We define $\mathcal{F}(w) \Equaldef \frac{\partial \tilde{f}_0}{\partial w}$ on $w\in(0,1)$. Therefore,
\begin{multline}\label{fp}
        \mathcal{F}(w) = \sum_{d=0}^N \binom{N}{d} \exp\bigg(\beta w \Big( \frac{d^2-d}{2}-\frac{\theta}{N}(N-1)d\Big)\bigg) \\
         \times \bigg(\beta \Big(\frac{d^2-d}{2}-\frac{\theta}{N}(N-1)d\Big)\bigg) +\frac{\rho}{2}N(N-1).
\end{multline}

We are interested in obtaining a sufficient condition for the existence of $w^\star \in (0,1)$ such that $\mathcal{F}(w^\star) = 0$.

\vspace{5pt}

\begin{theorem}\label{rtoc}
For any $\rho > 0$ and any $N$, there exists a sufficiently large $\beta$ such that $w^\star(\beta) \in (0,1)$. Moreover,
\begin{equation} 
\frac{\partial w^\star(\beta)}{\partial \beta} < 0.
\end{equation}
Therefore, increasing the rationality parameter $\beta$ leads to a decrease in the optimal agent connectivity.
\end{theorem}

\vspace{5pt}

\begin{proof} Notice that if $\theta>N/2$, the derivative $\mathcal{F}'(w)>0$, which implies that $\mathcal{F}$ is strictly monotone increasing.

To prove \cref{rtoc}, we need sufficient conditions for the existence of a $w^\star \in (0,1)$. For that, we use Intermediate Value Theorem to show that if $\mathcal{F}(0) <0$ and $\mathcal{F}(1)>0$, there must exist a unique  solution $w^\star\in (0,1)$ such that $\mathcal{F}(w^\star) = 0$.

Using \cref{fp}, the condition $\mathcal{F}(0) <0$ holds if and only if
    \begin{multline}
        \beta \Big[ \frac{N}{2}(N-1)2^{N-2}-N\Big(\frac{1}{2}+\frac{\theta}{N}(N-1)\Big)2^{N-1}\Big] \\
        < -\frac{\rho}{2}N(N-1)
    \end{multline}
    Solving for $\beta$, we get
    \begin{equation}
        \beta > \frac{2^{2-N}\rho N}{4\theta-N}.
    \end{equation}

Similarly, using \cref{fp},     the condition $\mathcal{F}(1)>0$ holds if and only if
\begin{multline}\label{eq:ineq_existence}
        \sum_{d=0}^N \binom{N}{d} \exp\bigg(\beta d \Big(\frac{d-1}{2}-\frac{\theta}{N}(N-1)\Big)\bigg)\\ \times \bigg(\beta d\Big(\frac{d-1}{2}-\frac{\theta}{N}(N-1)\Big)\bigg)>-\frac{\rho}{2}N(N-1).
    \end{multline}

 Note that for $\theta>N/2$, we have 
    \begin{equation}
     \frac{d-1}{2}-\frac{\theta(N-1)}{N} < 0,
    \end{equation}
    for all $d\in \{0,\ldots,N\}$.
   We will approximate the left hand side (LHS) of \cref{eq:ineq_existence} in the high rationality regime ($\beta \gg 0$). Define
    \begin{multline}
       \mathrm{LHS} \Equaldef \sum_{d=0}^N 
       \binom{N}{d}\exp\bigg(\beta d\Big(\frac{d-1}{2}-\frac{\theta}{N}(N-1)\Big)\bigg)\\ \times \bigg(\beta d\Big(\frac{d-1}{2}-\frac{\theta}{N}(N-1)d\Big)\bigg).
    \end{multline}
Then, for $\beta$ sufficiently large the summation in LHS is dominated by its largest term, i.e., 
\begin{multline}
        \mathrm{LHS} \approx 
     \binom{N}{d^\star}   \exp\bigg(\beta d^\star\Big(\frac{d^\star-1}{2}-\frac{\theta}{N}(N-1)\Big)\bigg)\\ 
    \times \bigg(\beta d^\star\Big(\frac{d^\star-1}{2}-\frac{\theta}{N}(N-1)\Big)\bigg),
    \end{multline}
    where $d^\star = \lceil \frac{\theta(N-1)}{N}+\frac{1}{2} \rceil $ is the value of $d\in \{0,\ldots,N\}$ that maximizes the argument of the exponential term. 
    Notice that even at $d = d^\star$, the exponent is negative, which implies that 
    in the limit of $\beta\rightarrow \infty$,  $\mathrm{LHS} \uparrow 0$. Therefore, there exists a sufficiently large $\beta$ such that $\mathcal{F}(1)>0$.
    Since $\mathcal{F}(0)<0$ establishes a lower bound on the rationality $\beta$, there always exist a large enough $\bar{\beta}$ such that both conditions will be satisfied for $\beta\geq \bar{\beta}$. Therefore, there must exist $w^\star(\beta) \in (0,1)$ such that $\mathcal{F}\big(w^\star(\beta)\big) = 0$.

\begin{figure*}[t]
    \centering
    \includegraphics[width=\textwidth]{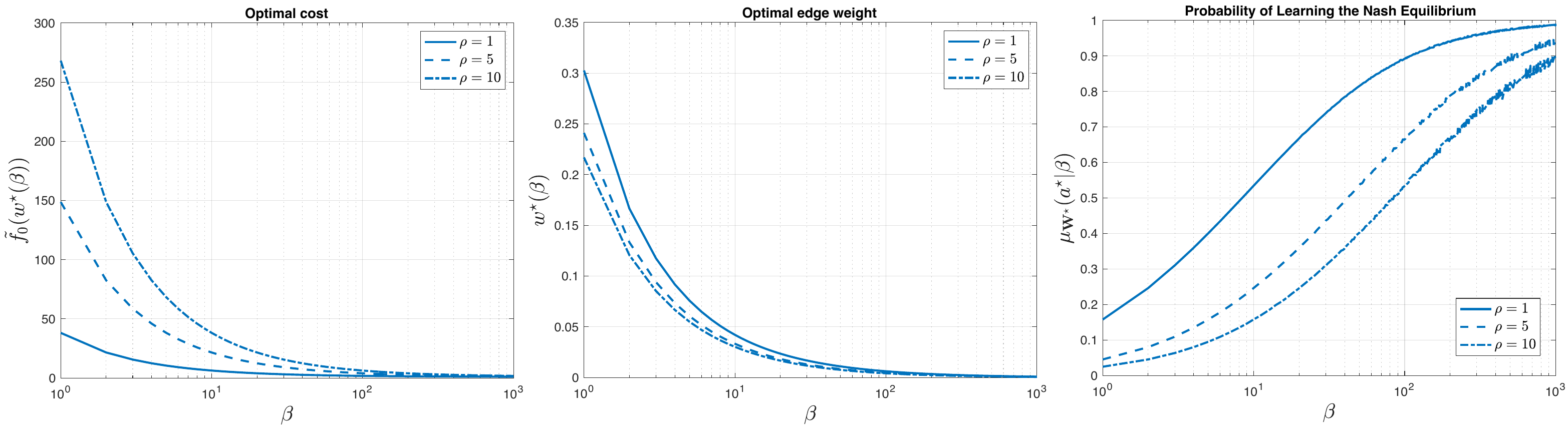}
    \caption{Numerical results for the optimal graph weight when $N=20$, $\theta=N/2+1$, and $\rho=1,5,10$ as a function of the rationality $\beta$: (left) shows the total optimal cost, (middle) shows the optimal edge weight, and (right) shows the resulting probability of playing the Nash equilibrium $a^\star=\mathbb{0}$.}
    \label{fig:my_two_column_figure}
\end{figure*}
    
  Finally, using the Implicit Function Theorem\footnote{Implicit Function Theorem: Since $\mathcal{F}(w^\star(\beta))=0$, the following holds
$$\frac{\partial w^\star}{\partial \beta}  = -\frac{\partial \mathcal{F}}{\partial \beta}\Big/\frac{\partial \mathcal{F}}{\partial w^\star}
$$}, we can compute the partial derivative of $w^\star(\beta)$ as follows
\begin{equation}\label{eq:derivative}
\frac{\partial w^\star(\beta)}{\partial \beta} = 
- \frac{
\sum\limits_{d=0}^N \binom{N}{d} \exp\left(\beta w^\star(\beta) A_d\right) \cdot \big(A_d + \beta w^\star(\beta)A_d^2\big)  
}{\beta^2 
\sum\limits_{d=0}^N \binom{N}{d} \exp\left(\beta w^\star(\beta) A_d\right) \cdot A_d^2
},
\end{equation}
where
\begin{equation}
A_d = \left( \frac{d^2 - d}{2} - \frac{\theta}{N}(N - 1)d \right).
\end{equation}

In the limit of $\beta\rightarrow \infty$, \cref{eq:derivative} can be approximated as
\begin{equation}
\frac{\partial w^\star(\beta)}{\partial \beta} \approx 
- \frac{w^\star(\beta)}{\beta} < 0.
\end{equation}

Since $w^\star(\beta) \in (0,1)$, there exists a large enough $\bar{\beta}$ such that  $\frac{\partial w^\star}{\partial \beta} < 0$ for all $\beta>\bar{\beta}$. Therefore, if the weight $w^\star \in (0,1)$, in optimality, rationality and connectivity are negatively correlated in the sense that more rational agents require less connectivity and are less influenced by other agents in the system.

\end{proof}

\subsection{Illustrative Example}

To illustrate our theoretical results, we consider the optimal design of the weighted graph for an unrestricted sparsity pattern. In other words, any pair of agents can be connected. Theorem 1 implies that there exists a fully connected graph with the uniform edge weight $w^\star$ that is optimal for the problem in \cref{eq:opt_prob}. Considering a system with $N$ agents and difficulty parameter $\theta=N/2+1$, the Nash equilibrium is $a^\star=\mathbb{0}$ (c.f. \cref{prop:nash_equilibria}). We find the optimal weight by computing $w^\star(\beta)\in (0,1]$ that minimizes \cref{eq:simplified_problem}.  

\Cref{fig:my_two_column_figure} shows the numerical results\footnote{The code and datasets used in this work are available at
 \url{https://github.com/MINDS-code/Teaming.git}} for $N=20$ and connectivity cost $\rho=1,5,10$ as a function of the agents' rationality $\beta$. In \cref{fig:my_two_column_figure} (left), we observe how the optimal cost decreases as a function of the rationality. In \cref{fig:my_two_column_figure} (middle), we observe how the optimal edge weight decreases as a function of $\beta$, which illustrates Theorem 2, showing that for sufficiently rational agents, the optimal weight decreases as the agents become more rational. Notice that as long as $\beta<\infty$, $w^\star(\beta)>0$, otherwise the algebraic connectivity constraint in \cref{eq:opt_prob} would be violated. 

Finally, in \cref{fig:my_two_column_figure} (right), we observe just the resulting stationary probability of agents playing the Nash equilibrium $a^\star=\mathbb{0}$ as a function of their rationality $\beta$. We see that the probability increases monotonically with $\beta$, but since the agents have bounded rationality, this probability will be bounded away from $1$, i.e., perfect collective learning cannot be achieved. However, our results stablish the maximum probability of playing a Nash equilibrium for a given level of connectivity that the system designer can afford.

\section{Extensions}

\subsection{Arbitrary sparsity patterns}

Our previous results hold for a possibly complete (i.e. unrestricted graph) where any two agents can be connected. However, the convexity of the optimization problem holds even when we constrain the graphs to have an arbitrary sparsity pattern, $\mathcal{S}$. The sparsity pattern of a graph specifies the connections/weights that must be equal to zero. As an example, the sparsity pattern for a weight matrix of a \textit{star network} with $N$ nodes is given by
\begin{equation}
\mathcal{S}_{\mathrm{star}} = \begin{bmatrix}
0 & * & * & \cdots & * \\
* & 0 & 0 & \cdots & 0 \\
* & 0 & 0 & \cdots & 0 \\
\vdots & \vdots & \vdots & \ddots & \vdots \\
* & 0 & 0 & \cdots & 0
\end{bmatrix} 
\end{equation}
Here, the first row and first column correspond to the center agent, which is adjacent to all other agents, and all other entries are zero, indicating no direct connection among the peripheral nodes. 

Similarly, the sparsity pattern for a weight matrix of a \textit{line network} with $N$ nodes is given by
\begin{equation}
\mathcal{S}_{\mathrm{line}}= \begin{bmatrix}
0    & *   & 0    & \cdots & 0 \\
*    & 0   & *    & \cdots & 0 \\
0    & *   & 0    & \ddots & \vdots \\
\vdots & \vdots & \ddots & \ddots & * \\
0    & 0   & \cdots & *    & 0 \\
\end{bmatrix}
\end{equation}
where the entry in the $i$-th row and $i+1$-th column (and symmetrically the $i+1$-th row and $i$-th column) is $*$, representing that only connections between consecutive nodes is possible.

With this additional constraint, we have the following convex optimization problem
\begin{equation}\label{eq:opt_prob_sparsity}
\begin{aligned}
\mathrm{minimize}  \quad & 
\frac{1}{\mu_\mathbf{W}(a^\star \mid \beta)}
+ \frac{\rho}{2}\mathbb{1}^\mathsf{T}\mathbf{W}\mathbb{1} \\
\text{subject to} \quad 
& \operatorname{trace}(\mathbf{W}) = 0 \\
& 0 \leq \mathbf{W} \leq 1 \\
& \mathbf{W} = \mathbf{W}^\mathsf{T} \\
& \lambda_2\big(\mathbf{L}(\mathbf{W})\big) > 0 \\
& \mathbf{W} \in \mathcal{S}
\end{aligned}
\end{equation}
with variable $\mathbf{W}\in \mathbb{R}^{N\times N}.$ 

\vspace{5pt}

\begin{remark}
To ensure at least one feasible solution exists, the sparsity pattern must be symmetric and allow at least one connected graph. Note that the sparsity pattern constraint is convex, and it differs from a constraint of the form “$\mathbf{W}$ corresponds to a star network,” since the convex combination of two distinct star networks is generally not a star network. However, the set of all star networks with a fixed central node and peripheral nodes is convex.
\end{remark}

\vspace{5pt}

\begin{figure}[t!]
    \centering
\includegraphics[width=0.6\linewidth]{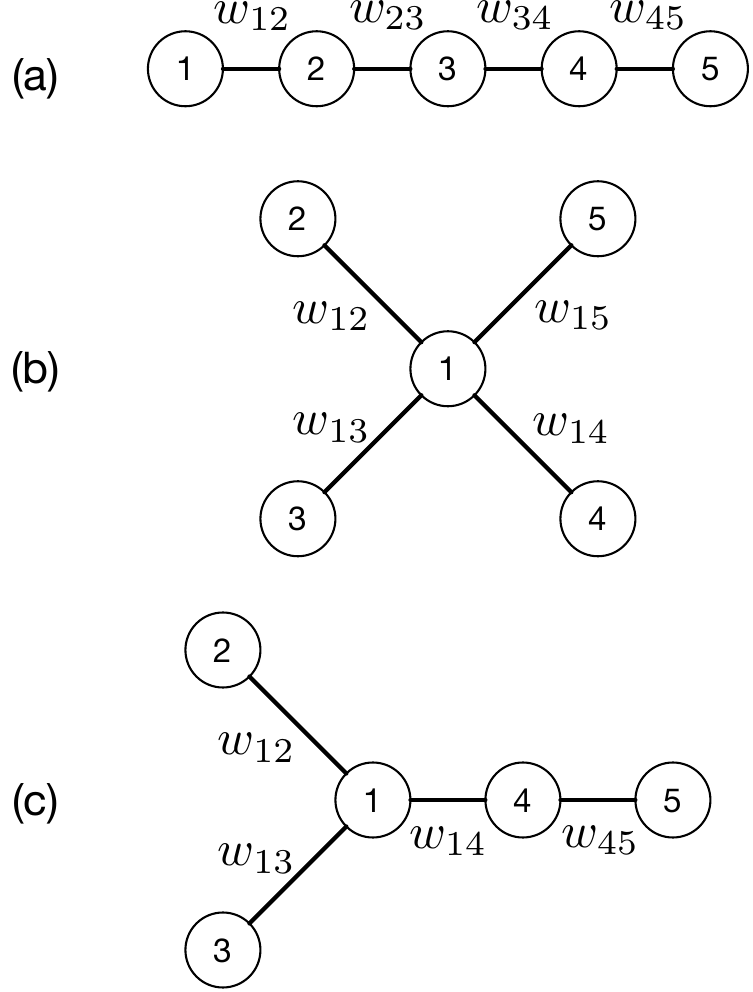}
    \caption{Examples of Networks with Incomplete Sparsity Patterns.}
    \label{fig:networks}
\end{figure}

To obtain the optimal weight matrix for networks with incomplete sparsity patterns, we consider the three networks shown in \cref{fig:networks}: a line network $(a)$, and star network $(b)$ and a hybrid between line and star network $(c)$. Notice that $(a),(b),(c)$ are networks with the same number of nodes, and the same number of edges in their respective sparsity patterns, which enables us to make a fair comparison between them.

The first observation is that for the star network $(b)$, the optimal solution is always symmetric. For instance, when $\beta=1$, $\theta=3$ and $\rho=5$, we have
\begin{equation}
\mathbf{W}_{\mathrm{star}}^\star =
\begin{bmatrix}
0 & 0.5174 & 0.5174 & 0.5174 & 0.5174 \\
0.5174 & 0 & 0 & 0 & 0 \\
0.5174 & 0 & 0 & 0 & 0 \\
0.5174 & 0 & 0 & 0 & 0 \\
0.5174 & 0 & 0 & 0 & 0
\end{bmatrix}
\end{equation}
with 
\begin{equation}
f_{0,\mathrm{star}}^\star = 26.5133.
\end{equation}

However, when we optimize the line network in $(a)$ for $\beta=1$, $\theta=3$ and $\rho=5$, we obtain 
\begin{equation}
\mathbf{W}_{\mathrm{line}}^\star =
\begin{bmatrix}
0 & 0.5363 & 0 & 0 & 0 \\
0.5363 & 0 & 0.5056 & 0 & 0 \\
0 & 0.5056 & 0 & 0.5056 & 0 \\
0 & 0 & 0.5056 & 0 & 0.5363 \\
0 & 0 & 0 & 0.5363 & 0
\end{bmatrix}
\end{equation}
with
\begin{equation}
f_{0,\mathrm{line}}^\star = 26.4753.
\end{equation}
Notice that $\mathbf{W}_{\mathrm{line}}^\star$ is no longer symmetric (i.e., the first and the last edges have higher weights that the edges in the middle). If we attempt to symmetrize the solution, we obtain
\begin{equation}
\mathbf{W}^{\mathrm{sym}}_{\mathrm{line}} =
\begin{bmatrix}
0 & 0.5210 & 0 & 0 & 0 \\
0.5210 & 0 & 0.5210 & 0 & 0 \\
0 & 0.5210 & 0 & 0.5210 & 0 \\
0 & 0 & 0.5210 & 0 & 0.5210 \\
0 & 0 & 0 & 0.5210 & 0
\end{bmatrix}
\end{equation}
with
\begin{equation}
f_{0}(\mathbf{W}^{\mathrm{sym}}_{\mathrm{line}})= 29.5182,
\end{equation}
which is suboptimal. Therefore, the symmetrization argument used in the proof of Theorem 1 does not hold when a line sparsity pattern is imposed as a constraint. It remains an open question which sparsity patterns admit symmetric optimal solutions. One such example where symmetric solutions are optimal is the star network.

\begin{figure}[t!]
    \centering
    \includegraphics[width=0.95\linewidth]{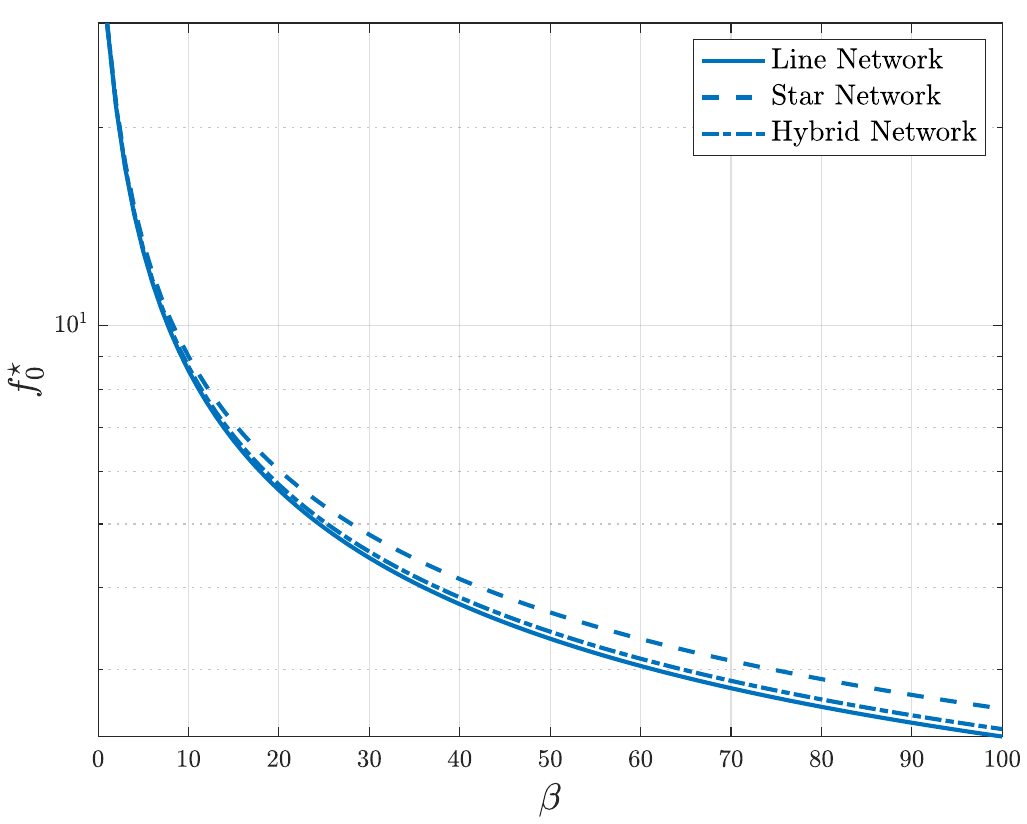}
    \caption{Optimized objective function value as a function of rationality parameter $\beta$ for three graph topologies.}
    \label{fig:objective_vs_beta}
\end{figure}

\subsection{Additional Numeric Results}

Lastly, we investigate the performance of the three networks $(a),(b),(c)$ as a function on $\beta$ for $\theta=4$ and $\rho=5$. Solving the optimization problem in \cref{eq:opt_prob_sparsity}, we obtain the following performance curves in \cref{fig:objective_vs_beta}, which shows the optimal objective value for each of the networks in \cref{fig:networks} as a function of the agents' rationality $\beta$. Although the overall performance of the three networks are very close, we can clearly see that the star network performs worse than the other two. Intuitively, in a star network, having the central agent be both highly connected and having the same bounded rationality as peripheral agents can hurt the overall team performance. 

It remains to compare the hybrid and line networks. Our numerical results in \cref{fig:objective_vs_beta} suggest that being connected in a line network is uniformly better than the hybrid one. This provides evidence that even in small-scale settings, there are team structures that are better than others. We conjecture that such performance improvement comes from the fact that the sparsity pattern $\mathcal{S}_{\mathrm{line}}$ is ``closer'' to the one of a regular graph. As established in \cite{zhang2024role}, when the graphs are unweighted, regular graphs maximize the probability of learning to coordinate at the Nash equilibrium. A rigorous analysis of this observation is left for future work.

\section{Conclusions}

We have addressed the problem of designing the connectivity graph that maximizes the probability of converging to an efficient Nash equilibrium of a binary network coordination game while minimizing the communication cost. We show that this is a convex optimization problem, but its solution is non-trivial since evaluating the objective function involves a sum over the entire action profile space, which grows exponentially with the number of agents in the network. Exploiting the underlying symmetry of the function, we prove that an optimal graph exists where the edges have uniform weights, which significantly reduces the computation of the optimal weight. For complete graphs, we showed that for sufficiently rational agents the optimal weight of the uniform optimal graph is monotone decreasing with respect to the rationality of the agents, which means that when agents are more rational, they can afford to be less connected to other agents in the network. Finally, we consider the optimization over different network sparsity patterns with the same number of nodes and edges and show that there are teaming structures that are better than others for coordination under bounded rationality. 

\bibliography{./reference/ref}
\bibliographystyle{ieeetr}

\end{document}